\documentclass[12pt]{article}
\usepackage{hyperref}

\begin{document}

\begin{center}

{\large ENERGY IN NEWTONIAN GRAVITY}

\vspace{10mm}

{\large Ingemar Bengtsson}

\vspace{7mm}

{\large Tobias Eklund}

\vspace{7mm}

{\sl Stockholms Universitet, AlbaNova\\
Fysikum\\
S-106 91 Stockholm, Sverige}

\vspace{7mm}

{\bf Abstract:}

\end{center}

{\small 

\noindent In Newtonian gravity it is a moot question whether energy should be localized in 
the field or inside matter. An argument from relativity suggests a compromise in which 
the contribution from the field in vacuum is positive definite. We show that the same 
compromise is implied by Noether's theorem applied to a variational principle 
for perfect fluids, if we assume Dirichlet boundary conditions on the potential. We 
then analyse a thought experiment due to Bondi and McCrea that gives a clean example of 
inductive energy transfer by gravity. Some history of the problem is included.}

\vspace{12mm}

{\bf 1. Introduction}

\vspace{5mm}

\noindent How is a local energy density to be defined in Newtonian gravity? This is a 
moot question. Traditionally, the two 
main contenders for a definition are that of Maxwell \cite{Maxwell}, 

\begin{equation} {\cal E}^\prime = - \frac{1}{8\pi G}\partial_i\Phi \partial_i\Phi \ , 
\end{equation}

\noindent and an alternative where the energy is localized within the matter, 

\begin{equation} {\cal E}^{\prime \prime} = - \frac{1}{2}\rho \Phi \ . \end{equation}

\noindent We will refer to the latter as Bondi's energy density, because it 
was championed by him \cite{Bondi, Morgan}. Our conventions for the gravitational 
interaction are set by the equations that connect the gravitational potential $\Phi$ 
to the mass density of matter $\rho$, 

\begin{equation} \nabla^2\Phi = - 4\pi G\rho \hspace{5mm} \Leftrightarrow 
\hspace{5mm} \Phi ({\bf x},t) = G \int \frac{\rho ({\bf x}',t)}{|{\bf x}-{\bf x}'|}
{\rm d}^3x' \ . \label{Poisson} \end{equation}

\noindent We have adopted the convention that the potential $\Phi$ is positive. At 
some points we may set the constant $G$ to 1. The formula using the Green function 
assumes that the matter density has compact support, so that we deal with an 
isolated system. 

The total energy is obtained by integrating the energy density over all space, and 
comes out the same regardless of whether we use ${\cal E}^\prime$ or 
${\cal E}^{\prime \prime}$. There is an analogous issue in electrostatics. 
Maxwell, who was thinking of force fields as emergent from a medium, insisted that 
the choice is important: 

\

\noindent {\small ``I wish to be understood literally. All energy 
is the same as the mechanical energy, whether it exists in the form of motion or in that 
of elasticity \ \dots The only question is, Where does it reside? ... On our theory it resides 
in the electromagnetic field, in the space surrounding the electrified and magnetic 
bodies, as well as in those bodies themselves.'' \cite{Maxwell}}

\

\noindent But Maxwell faced a problem with gravity. The energy density 
is negative definite while 
(in his view) energy is ``essentially positive''. As a way out he suggested that a 
constant should be added to ${\cal E}^\prime$ to make it everywhere positive. But 
this would mean that the energy density of the medium must be huge where the gravitational 
field is weak. He concluded: 

\

\noindent {\small ``As I am unable to understand in what way a medium can possess such 
properties, I cannot go any further in this direction in searching for the cause of 
gravitation.''~\cite{Maxwell}}

\ 

\noindent It should be noticed that the notion of energy conservation was by no 
means uncontroversial at the time. Herschel regarded it as a verbal trick \cite{Herschel}. 
Much later, Mason and Weaver \cite{Mason} argued that energy is a 
function of the configuration of the system as a whole, and that it is no 
more sensible to inquire about the location of energy than to declare that the 
beauty of a painting is distributed over the canvas in a specified manner. 

Who is right: Maxwell, Bondi, or Mason and Weaver? For electromagnetism the question 
is often regarded as resolved (in favour of Maxwell) by the relativistic theory, and 
in particular by the way that the electromagnetic field couples to gravity. For gravity 
there is no external arbiter to make the decision. 

In general relativity the total energy of an isolated system is well understood, but 
the localisation of energy is a tangled question indeed. 
The energy at a point can be argued away using the equivalence principle, but there 
are proposals for the energy located within some chosen closed surface. Some of these 
proposals, notably those of Hawking \cite{Hawking} and Penrose \cite{Penrose}, suggest 
that the energy of a black hole is located within its event horizon. This is perhaps 
reminiscent of Bondi's expression in the Newtonian case. Others, notably Lynden-Bell 
and Katz \cite{LyndenBell}, have proposed expressions that are closer in spirit to that 
of Maxwell. To support their case Lynden-Bell and his coworkers considered a static 
gravitational field coupled to a perfect 
fluid. We then lose very little by assuming spherical symmetry as well, so that the 
formulas that follow should be familiar to most readers. To obtain a conserved current 
from the relativistic stress-energy tensor $T_{ab}$ we need to contract it with a timelike 
Killing vector $\xi^b$, and the energy density 
gains an extra factor coming from the norm of 
the Killing vector field. Let $t^a$ be a timelike unit vector and set 

\begin{equation} T_{ab} = (\rho + p)t_at_b + pg_{ab} \end{equation} 

\begin{equation} \xi^a = \sqrt{-g_{tt}}t^a \ , \hspace{5mm} - g_{tt} = 1 -2\Phi \ . 
\end{equation} 

\noindent We can build an energy density $\mu$ and take its Newtonian limit, 

\begin{equation} \mu = - T_{tb}\xi^b = \rho \sqrt{1 - 2\Phi} \approx 
\rho (1 - \Phi ) = \rho + 2{\cal E}^{\prime \prime} \ . \end{equation}

\noindent Maxwell's very large constant appears here in the guise of the mass density 
$\rho$, but the binding energy is twice as large as he may have expected. To make the 
total energy come out as being equal to the mass plus the Newtonian energy we add a 
term coming from the gravitational field itself. We can then define a `true' Newtonian 
energy density ${\cal E}$ through \cite{LyndenBell, HorStro} 

\begin{equation} \rho + {\cal E} = \rho + 2{\cal E}^{\prime \prime} - {\cal E}^\prime = 
\rho(1 - \Phi ) + \frac{1}{8\pi}\partial_i\Phi \partial_i\Phi \ . 
\label{Horowitz} \end{equation}

\noindent Maxwell's sign problem has evaporated in the sense that the energy density is now 
positive definite outside matter, although the argument rests on the 
assumption that the gravitational field is static. 

This argument did not lead to a consensus. Frauendiener and Szabados \cite{FS} use post-Newtonian 
corrections to Newton's theory to argue that the energy density integrated over large 
volumes, enclosing all the matter, should be a monotonically decreasing function of the volume. 
So the question is indeed moot.

As far as we know the expression ${\cal E} =2 {\cal E}^{\prime \prime} - {\cal E}^\prime$ 
for the energy density was first advocated by Ohanian \cite{Ohanian}. The question continues to attract 
interest \cite{Sebens}. Our first aim here is to see how one can argue in favour of the energy density 
${\cal E}$ from within the Newtonian theory itself, at the same time dropping the assumption of static 
fields. We will do this by appealing to Noether's theorem. In view of its hundreth 
anniversary this theorem has attracted interest from philosophers of science recently 
\cite{DeHaro}, and indeed Dewar and Weatherall used it to study the energy concept 
in Newtonian gravity \cite{DW}. However, since they used an external matter source 
they were unable to address the question that we consider in Section 2. We add that other 
aspects of their paper have given rise to illuminating discussions \cite{Read}. In particular 
the Newton--Cartan formulation of the theory has been a subject for these discussions. 
This is an interesting topic about which we have nothing to say. 

In Section 3 we go on to consider an interesting example of energy transport in 
Newtonian gravity, in rather more detail than was offered in the original paper by 
Bondi and McCrea \cite{BondiMcC}. In Section 4 we draw attention to the fact that 
the energy density proposed in Section 2 plays no role in the concrete setting 
of Section 3. The question then becomes one of the pragmatic advantages of the 
various definitions. 

\vspace{1cm}

{\bf 2. An energy density from Noether's theorem}

\vspace{5mm}

\noindent In Newtonian gravity we need matter to provide dynamics. A fluid described 
by a mass density $\rho$ and a velocity field $v_i$ is appropriate. These variables 
obey mass conservation 

\begin{equation} \partial_t\rho + \partial_i(\rho v_i) = 0 \label{masskonserv} \end{equation}

\noindent as well as the Euler equation 

\begin{equation} \rho\frac{\rm d}{{\rm d}t}v_i = \rho(\partial_t v_i + v_j\partial_jv_i) 
= \partial_j\tau_{ij} + \rho \partial_i\Phi \ . \label{Euler} \end{equation}

\noindent Here $\tau_{ij}$ is the stress tensor, and the last term is the gravitational 
body force. In the spirit of Maxwell the latter can be regarded as being due to 
gravitational stress, but for this we refer to Synge \cite{Synge}. 
For simplicity we will set $\tau_{ij}$ to zero and assume that the velocity field is 
irrotational. Thus our matter consists of irrotational gravitating dust, but we comment 
briefly on more general cases at the end. That the velocity field is irrotational means 
that there exists a velocity potential $\lambda$, 

\begin{equation} \partial_iv_j - \partial_jv_i = 0 \hspace{5mm} \Leftrightarrow 
\hspace{5mm} v_i = - \partial_i\lambda \ . \label{irrot} \end{equation}

\noindent The potential is essential in order to find an action integral from which 
Euler's equations can be derived \cite{Seliger, SS, Mukunda}. 

We begin by recalling how the variational principle works. In the action the velocity 
potential appears 
as a Lagrange multiplier imposing conservation of mass: 

\begin{equation} S_0[\rho, v_i, \lambda, \Phi] = \int \left[ \frac{1}{2}\rho v_iv_i - 
\lambda \left( \partial_t\rho + \partial_i(\rho v_i)\right) + \rho \Phi 
- \frac{1}{8\pi}\partial_i\Phi \partial_i\Phi \right] {\rm d}^4x \ . \end{equation}

\noindent Varying the action with respect to the velocity field $v_i$ we recover 
equation (\ref{irrot}), which is an algebraic equation for $v_i$ that can be inserted 
in the action. The result is 

\begin{equation} S_1[\rho, \lambda , \Phi] = \int \left[ \rho \partial_t\lambda - 
\frac{1}{2}\rho \partial_i\lambda \partial_i\lambda + \rho \Phi - 
\frac{1}{8\pi}\partial_i\Phi \partial_i\Phi  \right] {\rm d}^4x\ . \label{action1} \end{equation}

\noindent A total time derivative was added. This amounts to a canonical transformation 
in the matter sector, and does not affect our later arguments. Total space derivatives are 
important and will be discussed soon. Variation with respect to 
$\Phi$ returns Poisson's equation (\ref{Poisson}), and variation with respect to $\rho$ gives 

\begin{equation} \partial_t\lambda - \frac{1}{2} \partial_i\lambda \partial_i\lambda + 
\Phi = 0 \ . \end{equation}

\noindent Taking the gradient of this equation and making use of eq. (\ref{irrot}) yields 

\begin{equation} - \partial_t\partial_i\lambda + \partial_j\lambda \partial_j\partial_i 
\lambda = \partial_i\Phi \hspace{5mm} \Rightarrow \hspace{5mm} 
\partial_tv_i + v_j\partial_jv_i = \partial_i \Phi \ . \end{equation}

\noindent This is the equation of motion for irrotational dust. 

The velocity potential $\lambda$ has no direct physical interpretation since it 
is defined only up to a constant. In a Galilei invariant model we insist on 
the mass superselection rule, that is we insist that all observables Poisson 
commute with the total mass \cite{Mukunda} 

\begin{equation} M = \int \rho \ {\rm d}^3x \ . \end{equation}

\noindent Clearly $\{ M, \lambda\} = 1$, so indeed $\lambda$ is not an observable while 
its gradient is. 

Before we apply Noether's theorem we must address the issue of possible surface terms 
that can be added to the action. In infinite space they play no role in deriving the equations 
of motion, but we will consider a finite spacetime region $V$ with an enclosing surface placed in vacuum 
\cite{Kijowski}. The action amended with a surface term is 

\begin{eqnarray} S[\rho, \lambda , \Phi] = S_1[\rho, \lambda , \Phi]  + \frac{a}{4\pi} \int 
\partial_i (\Phi \partial_i\Phi ) {\rm d}^4x  = \hspace{30mm} \nonumber \\ \label{action} \\ 
= \int \left[ \rho \partial_t\lambda - 
\frac{1}{2}\rho \partial_i\lambda \partial_i\lambda + \rho \Phi - 
\frac{1}{8\pi}\partial_i\Phi \partial_i\Phi + 
\frac{a}{4\pi}\partial_i (\Phi \partial_i\Phi ) \right] {\rm d}^4x \ .  \nonumber \end{eqnarray}

\noindent Here $a$ is a parameter to be fixed. Assuming that the field equations hold we find that 

\begin{equation} \delta S = \int_V \left[ \partial_t (\delta \lambda \rho ) + \partial_i 
\left( \rho v_i \delta \lambda + \frac{a-1}{4\pi}\partial_i \Phi \delta\Phi + 
\frac{a}{4\pi}\Phi \partial_i\delta \Phi 
\right) \right] {\rm d}^4x \ . \end{equation}

\noindent The surface terms must vanish if the action is to be used to derive the field equations in the bounded region. 
The total time derivative is not relevant for our problem, but the total divergence is. We have assumed that $\rho = 0$ 
on the spatial part of the boundary, but we still have to decide what boundary conditions to impose 
on $\Phi$. This is the question of how we control the system \cite{Kijowski}, and an answer will fix the 
parameter $a$. The two most obvious choices give 

\begin{equation} \begin{array}{ccccc} \mbox{Dirichlet} & \Rightarrow & \delta \Phi_{\rm bdry} = 0 & \Rightarrow & a = 0 \ \ \\ 
\\ \mbox{Neumann} & \Rightarrow &n^i \delta \partial_i\Phi_{\rm bdry} = 0 & \Rightarrow & a = 1 \  , \end{array}
\end{equation}

\noindent where  $n^i$ is normal to the boundary. In electrostatics it would be a straightforward task for the experimentalist 
to impose Dirichlet conditions on the electrostatic potential.  For gravity the question is not so easily settled, but Dirichlet conditions 
do seem to be a natural choice. We will return to discuss this issue once we have discussed a concrete example of gravitational 
energy transport. Meanwhile we simply assume that a suitable choice has been made, so that the parameter $a$ is 
fixed. 

We are now ready to apply Noether's theorem. Consider a rigid time translation, 

\begin{equation} \delta \lambda = \epsilon \partial_t\lambda \hspace{4mm} \delta \rho = 
\epsilon \partial_t\rho \hspace{4mm} \delta \Phi = \epsilon \partial_t\Phi \hspace{5mm} 
\Rightarrow \hspace{5mm} \delta {\cal L} = \epsilon \partial_t\cal L \ , \end{equation}

\noindent where $\cal L$ is the integrand of the action integral. Noether's theorem 
follows from the observation that for these variations we have the alternative expression 

\begin{equation} \delta S 
= \int \partial_t(\epsilon {\cal L}) \ {\rm d}^4x\ . \end{equation}

\noindent The equality of the two expressions for $\delta S$ implies the 
local conservation law 

\begin{eqnarray} \partial_t \left( \frac{1}{2}\rho \partial_i\lambda \partial_i\lambda 
+ (a-1) \rho \Phi + \frac{1-2a}{8\pi}\partial_i\Phi \partial_i\Phi \right)  + \ \nonumber \\ 
\label{konserv} \\ 
+ \partial_i \left( \rho v_i\partial_t\lambda  + \frac{a-1}{4\pi}\partial_t\Phi \partial_i\Phi 
+ \frac{a}{4\pi}\Phi \partial_i\partial_t\Phi \right) = 0 \ . \nonumber \end{eqnarray}

\noindent The energy density (including the kinetic energy of matter) can be read off from the 
first term. If Dirichlet conditions are assumed we set $a = 0$ and find that the energy density 
is  

\begin{equation} {\cal E} = \frac{1}{2}\rho v_iv_i - \rho \Phi + 
\frac{1}{8\pi}\partial_i\Phi \partial_i\Phi = \frac{1}{2}\rho v_iv_i - {\cal E}^\prime 
+ 2{\cal E}^{\prime \prime} \ . \end{equation}

\noindent This is not a decision between Maxwell and Bondi, it is a compromise in the 
sense that energy is partly localized within matter and partly spread throughout the vacuum. 
It bears the mark of a good compromise since it agrees with the answer arrived at by 
Lynden-Bell and Katz, as in eq. (\ref{Horowitz}). And the point 
we wanted to make is precisely that with suitable assumptions this answer can be 
derived from within the Newtonian theory itself. 

However, had we chosen to impose Neumann rather than Dirichlet boundary conditions we would 
have ended up with Maxwell's expression for the energy density. That Noether's theorem 
leads to different expressions for the energy depending on the boundary conditions 
we impose should, in fact, not be surprising. The situation is interestingly analogous to 
how internal and free energy appear in thermodynamics, depending on how the 
system is controlled \cite{Kijowski}. 

The second term in the local conservation law (\ref{konserv}) gives the local energy flux. 
We will return to it in Section 4 below. Meanwhile we observe that if we use the field 
equations to clear the conservation law of time derivatives we obtain 

\begin{equation} \partial_t \left( \frac{1}{2}\rho v_iv_i\right) + \partial_i
\left( v_i \frac{1}{2}\rho v_jv_j\right) = \rho v_i\partial_i\Phi \ . \end{equation}

\noindent This is of course indisputable---the local kinetic energy changes due to work 
done by the gravitational field---but any gravitational contribution to the energy 
density has disappeared. 

We have given the argument for irrotational dust. The entire argument clearly goes 
through if we add a pressure term to the equations. The restriction to irrotational 
flow can be dropped too, if we make use of Clebsch potentials. 
An economical choice is that due to Seliger and Whitham \cite{Seliger}, 

\begin{equation} v_i = - \partial_i\lambda - \alpha \partial_i\beta \ . \end{equation}

\noindent An additional pair of Clebsch potentials is needed to handle general flows globally \cite{Bretherton}. 
Applying Noether's theorem to the action proposed by Seliger and 
Whitham results in the same expressions for the local energy density and the local energy 
transport as the ones we just derived, once they have been expressed in terms of 
$\rho$, $v_i$, and $\Phi$. For this reason we do not give the details here. 

\vspace{1cm}

{\bf 3. Tweedledum and Tweedledee}

\vspace{5mm}

\noindent The arguments in Section 2 were rather formal. It is interesting to see how they 
fare in a concrete problem. We do know that energy is being transported by 
gravity within the solar system, in a way that is well described by Newtonian theory. 
As an example, tidal friction within the Earth--Moon system causes the Moon to move away 
from the Earth. But in 
this case it is not easy to pinpoint where the energy ends up. A more dramatic 
example is provided by the tidal heating of Io, one of the moons of Jupiter \cite{Io}. 
A simpler example is called for here. Bondi and McCrea invented a thought experiment 
in which tidal forces give rise to a net energy 
transport even though the gravitational field returns to its initial state after some 
energy has been transmitted \cite{BondiMcC}. The experiment concerns two mutually 
gravitating bodies in elliptical orbits around each other. Bondi later named them 
Tweedledum and Tweedledee \cite{Bondi}, although in the original paper they were 
referred to as the receiver (R) and the transmitter (T). They have spherical outlines, 
but their mass distributions can be changed between 
prolate and oblate with the axial direction orthogonal to the orbital plane. If they 
both turn oblate the gravitational attraction between them grows. If one of them turns 
oblate and the other prolate the attraction can be kept constant, and this is the key 
to the whole idea since it will allow them to stay on their elliptical orbits throughout the 
duration of the experiment. The changes of shape are controlled by some machinery powered 
by batteries external to the system that we will describe, which goes to say that we will 
study an open system. Tidal forces tend to make a 
body oblate, and when this happens work is done on the body. Conversely, work is done by 
the body when it turns prolate. This is how transmission of energy can occur. The question 
of how energy is stored in the gravitational field is avoided because, as far as the 
gravitational field is concerned, the process is cyclic and we will compute the 
energy transmitted during a full cycle. 

The original paper is very brief, and we feel that it may be useful to tell the 
story using equations. Full calculational details are given elsewhere~\cite{Tobias}. 
The twins are modelled as spheres with radii $r_0$ and total mass $M$ in both cases. 
Their mass densities have time dependent quadrupole moments $Q_{\rm R} =Q_{\rm R}(t)$ 
and $Q_{\rm T} =Q_{\rm T}(t)$ that can be freely prescribed. In coordinate systems 
with origos at the centres of the bodies 

\begin{equation} \rho_{\rm R} = \left\{ \begin{array}{lcl} 
\frac{3}{4\pi}\frac{M}{r_0^3} + \frac{35}{4\pi}\frac{Q_{\rm R}(t)}{r_0^7} r^2
P_2(\cos{\theta}) & , & 
r < r_0 \ \\ 0 & , & r > r_0 \end{array} \right. \label{singlebody} \end{equation}

\begin{equation} \rho_{\rm T} = \left\{ \begin{array}{lcl} 
\frac{3}{4\pi}\frac{M}{r_0^3} + \frac{35}{4\pi}\frac{Q_{\rm T}(t)}{r_0^7} r^2
P_2(\cos{\theta}) & , & 
r < r_0 \ \\ 0 & , & r > r_0 \end{array} \right.  \end{equation}

\noindent where $P_2$ is a Legendre polynomial and $\theta$ is the angle against the 
normal to the orbital plane. A body is prolate if its quadrupole moment is positive. 

For the calculations to follow we need some formulas for the spherical harmonics 
$Y_{\ell,m}$. We recall that 

\begin{equation} P_\ell ({\bf x}) = \sqrt{\frac{4\pi}{2\ell + 1}}Y_{\ell,0}({\bf x}) \ ,
\end{equation}

\noindent where the functions depend only on the direction of the vector in the argument. 
The direction cosine of the vector is taken relative an axis orthogonal 
to the orbital plane. For vectors ${\bf x}$, ${\bf y}$ with lengths related by $y < x$ 
the translation theorem states that 

\begin{equation} \frac{Y_{\ell ,m}({\bf x}- {\bf y})}{|{\bf x}-{\bf y}|^{\ell +1}} = 
\sum_{\ell'}\sum_{m'}C_{\ell,\ell'}^{m,m'}\frac{y^{\ell'}}{x^{\ell+\ell'+1}}
Y_{\ell+\ell',m+m'}({\bf x})Y_{\ell',m'}^*({\bf y}) \end{equation}

\noindent where the expression for the coefficients $C_{\ell,\ell'}^{m,m'}$ is somewhat 
unwieldy. Because we are going to integrate against the monopole-quadrupole massdensities 
we need only two terms in each of two special cases, 

\begin{equation} \frac{1}{|{\bf x}-{\bf y}|} = \frac{1}{x} + \dots + 
\frac{y^2}{x^3}P_2({\bf x})P_2({\bf y}) + \dots \label{translation1} \end{equation} 

\begin{equation} \frac{P_2({\bf x}-{\bf y})}{|{\bf x} - {\bf y}|^3} = 
\frac{P_2({\bf x})}{x^3} + \dots + \frac{6y^2}{x^5}P_4({\bf x})P_2({\bf y}) + 
\dots \ . \label{translation2} \end{equation}

\noindent For a proof see van Gelderen, who has an easily corrected misprint in his 
expression for $C_{\ell,\ell'}^{m,m'}$ \cite{Gelderen}. 

Using equation (\ref{translation1}) we can now calculate the gravitational potential 
outside the receiver as 

\begin{equation} \Phi_{\rm R} ({\bf x},t) = G \int \frac{\rho_{\rm R} ({\bf x}',t)}{|{\bf x} 
-{\bf x}'|}{\rm d}^3x' = \frac{GM}{r} + \frac{GQ_{\rm R}}{r^3}P_2({\bf x}) 
\ . \end{equation}

\noindent There is a similar formula for $\Phi_{\rm T}$. We recall that, in Newtonian 
gravity, the total self-force on a body vanishes \cite{Dixon, Harte}, and we 
calculate the total work needed to place the receiver at a distance $D$ from 
the transmitter. Using a vector ${\bf D}$ of length $D$ we need to calculate 

\begin{equation} V_{\rm R} = - \int \rho_{\rm R}\Phi_{\rm T} 
\ {\rm d}^3x= - \int \rho_{\rm R}({\bf x})\Phi_{\rm T}({\bf x} + {\bf D}) 
\ {\rm d}^3x \ . \end{equation} 

\noindent Appealing to equation (\ref{translation2}) we find that 

\begin{equation} V_{\rm R} = - \frac{GM^2}{D} + \frac{GM(Q_{\rm R} + Q_{\rm T})}{2D^2} 
- \frac{9GQ_{\rm R}Q_{\rm T}}{4D^5} \ . \label{VR} \end{equation}

By the terms of the agreement between the twins Tweedledum can choose the 
quadrupole moment $Q_{\rm R} = Q_{\rm R} (t)$ at will, but Tweedledee must then adapt the 
function $Q_{\rm T} = Q_{\rm T} (t)$ in such a way that the gravitational force between them 
stays the same as in the two-body problem for spherical bodies. Thus we impose 

\begin{equation} \frac{\partial V_{\rm R}}{\partial D} = \frac{GM^2}{D^2}
- \frac{3GM(Q_{\rm R} + Q_{\rm T})}{2D^4} + \frac{45GQ_{\rm R}Q_{\rm T}}{4D^6}
= \frac{GM^2}{D^2} \ . \end{equation}

\noindent The solution is 

\begin{equation} Q_{\rm T} = - \frac{2MD^2Q_{\rm R}}{2MD^2-15 Q_{\rm R}} \ . 
\label{eta} \end{equation}

\noindent For later use we also record the differential form of the constraint, 
\begin{equation} (15Q_{\rm T}-2MD^2){\rm d}Q_{\rm R} + (15Q_{\rm R}-2MD^2){\rm d}Q_{\rm T} 
- 4MD(Q_{\rm R} + Q_{\rm T}){\rm d}D = 0 \ . \label{diff} \end{equation}

\noindent We assumed that $2MD^2 > 15Q_{\rm R}$. With this precise choice for the 
quadrupole moments the orbits of the twins 
are ellipses with a common focus at the midpoint of 
the line between them, and the mutual distance $D = D(t)$ has a specified time dependence. 
It is a periodic function, and we assume that Tweedledum chooses a $Q_{\rm R}$ with 
the same periodicity. 

We now wish to calculate the rate of work $\dot{W}_{\rm R}$ done on the receiver as his 
quadrupole moment is changing. Bondi and McCrea use an elegant shortcut for this purpose, 
but it is interesting to calculate it using field theory. For clarity we decide that 
$\dot{W}_{\rm R}$ is to be calculated in an inertial system where the centre of mass of 
the receiver is momentarily at rest, while $\dot{W}_{\rm T}$ is calculated in a 
system where the transmitter is momentarily at rest. 

From equation (\ref{Euler}) we see that the rate of gravitational work is 

\begin{equation} \dot{W}_{\rm R} = \int \rho_{\rm R}v_i\partial_i\Phi_{\rm T} \ 
{\rm d}^3x \ . \end{equation}

\noindent Making use of the conservation of mass, equation (\ref{masskonserv}), this 
can be rewritten in two useful ways. Either as 

\begin{equation} \dot{W}_{\rm R} = \int \left[ \partial_i(\rho_{\rm R} v_i \Phi_{\rm T}) - 
\Phi_{\rm T}\partial_i(\rho_{\rm R}v_i)\right] {\rm d}^3x = \int \Phi_{\rm T}\partial_t
\rho_{\rm R} \ {\rm d}^3x \end{equation}

\noindent where a surface term was discarded, or as 

\begin{equation} \dot{W}_{\rm R} = \int \rho_{\rm R}\left( \frac{\rm d}{{\rm d}t}\Phi_{\rm T} 
- \partial_t\Phi_{\rm T}\right) {\rm d}^3x = - \frac{\rm d}{{\rm d}t}V_{\rm R} - 
\int \rho_{\rm R}\partial_t\Phi_{\rm T} \ {\rm d}^3x \end{equation}

\noindent where we made use of the material time derivative, and again used conservation 
of mass in the second step. The first way is slightly objectionable since it assumes 
that the mass distribution is smooth, whereas in fact we have chosen it to be 
discontinuous. This can be repaired by providing the bodies with a smooth skin. 
There is no such objection to the second way, and we will see that the two ways 
of calculation give the same result. 

Pursuing the first way of calculation we note that in the chosen inertial system the 
time dependence in $\rho_{\rm R}$ enters only through the function $Q_{\rm R}(t)$. Using 
the translation theorem we find that 

\begin{eqnarray} \dot{W}_{\rm R} = \int \partial_t\rho_{\rm R}({\bf x}) 
\Phi_{\rm T}({\bf x} + {\bf D}) {\rm d}^3x = \hspace{45mm} \nonumber \\ \\ = 
- \frac{GM}{2}\frac{\dot{Q}_{\rm R}}{D^3} \left( 1 - \frac{9Q_{\rm T}}{2D^2}\right) 
= - GM \frac{MD^2 - 3Q_{\rm R}}
{2MD^2 - 15Q_{\rm R}}\frac{\dot{Q}_{\rm R}}{D^3} \ . \nonumber \end{eqnarray}

\noindent In the last step we used the constraint (\ref{eta}) between the quadrupole 
moments. Similarly 

\begin{equation} \dot{W}_{\rm T} = - GM \frac{MD^2 - 3Q_{\rm T}}
{2MD^2 - 15Q_{\rm T}}\frac{\dot{Q}_{\rm T}}{D^3} \ . \end{equation}

\noindent Using the constraint (\ref{diff}) we find (after some calculation) that 

\begin{equation} \dot{W}_{\rm R} + \dot{W}_{\rm T} = \dot{F} \ , \end{equation}

\noindent where 

\begin{equation} F = \frac{9GQ_{\rm R}Q_{\rm T}}{4D^5} - 
\frac{GM(Q_{\rm R} + Q_{\rm T})}{2D^2} = V_{\rm R} + \frac{GM^2}{D} \ . \end{equation}

\noindent If we integrate to find the total amount of work transmitted during a 
cycle we find 

\begin{equation} W_{\rm R} + W_{\rm T} = \oint {\rm d}F = 0 \ . \end{equation}

\noindent Hence all of the energy transmitted by Tweedledee is received by Tweedledum. 

For the second way of calculation it is convenient to choose an inertial system 
in which the transmitter is momentarily at rest. Then we are no longer calculating 
the same thing. Denoting the rate of work done on the receiver in an inertial 
system where the transmitter is momentarily at rest by $\dot{W}_{\rm}^\prime$, we find 

\begin{equation} \dot{W}_{\rm R}^\prime + \frac{\rm d}{{\rm d}t}V_{\rm R} = - 
\int \rho_{\rm R}({\bf x}) \partial_t\Phi_{\rm T}({\bf x} + {\bf D}) {\rm d}^3x = 
- \dot{W}_{\rm T} \ . 
\end{equation}

\noindent Comparing the equations obtained, and recalling expression (\ref{VR}) 
for $V_{\rm R}$, we see that they are fully consistent, once we observe that 

\begin{equation} \dot{W}_{\rm R}^\prime = \dot{W}_{\rm R} + \frac{\rm d}{{\rm d}t}\left( 
\frac{GM^2}{D}\right) \ . \end{equation}

\noindent The motion of the centre of mass explains the difference. 

How much energy is being transmitted? With $Q_{\rm R} = Q_{\rm R} (t)$ we get 

\begin{equation} W_{\rm R} = \oint \dot{W}_{\rm R} {\rm d}t = 
\frac{GM}{2}\oint \frac{2MD^2 - 6Q_{\rm R}}
{2MD^2 - 15Q_{\rm R}}\frac{dQ_{\rm R}}{D^3} \ . \label{arbetsintegral} \end{equation}

\noindent Tweedledum's aim is to maximize this expression. We see that it can be made 
positive by letting the function $Q_{\rm R}(t)$ lag behind the periodic function $D = 
D(t)$ so that its derivative is negative when $D$ is close to its minimum. and positive 
when $D$ is close to its maximum. This will ensure that the total amount of work on 
the receiver in a cycle is positive. A natural choice is to let both bodies be spherical 
at the moment of closest approach, and at the moment when they are at maximal distance 
from each other. 

From the solution of the two-body problem we know that 

\begin{equation} D = D(\varphi (t)) = \frac{D_{\rm min}(1+e)}{1+e\cos{\varphi}} \end{equation}

\noindent where $e$ is the eccentricity of the ellipses and $\varphi$ is 
an angular coordinate with respect to the centre of mass. For definiteness we let 
Tweedledum choose 

\begin{equation} Q_{\rm R} = - Q_{\rm R}^{\rm max}\sin{\varphi} \ . \end{equation}

\noindent The work integral (\ref{arbetsintegral}) is easily approximated if the bodies 
are small compared to the distance between them, that is if $Q_{\rm R} << MD^2$. Then 
we obtain the positive result 

\begin{equation} W_{\rm R} \approx - \frac{GM}{2}\oint \frac{{\rm d}Q_{\rm R}}{D^3} = 
\frac{3\pi}{8}\frac{e(4+e^2)}{(1+e)^3}\frac{GMQ_{\rm R}^{\rm max}}{D_{\rm min}^3} \ . 
\end{equation}

\noindent The function of $e$ that occurs here has a maximum at $e = 2/3$. 

The full work integral (\ref{arbetsintegral}) is best treated numerically. 
It can be seen that it is always positive and that it diverges as the limit 
$2D^2M = 15Q_{\rm R}$ is approached \cite{Tobias}. Thus it is clear that the twins 
can achieve their aim of transmitting energy from the one to the other through 
the gravitational field. 

\vspace{1cm}

{\bf 4. Energy transfer}

\vspace{5mm} 

\noindent Reading Bondi and McCrea behind a veil of hindsight, and knowing that no-one did more 
than Bondi to prove that gravitational waves are real and do carry energy away, it 
is easy to read their paper as an argument for the reality of gravitational waves. This is 
probably a misreading of history though, since Bondi approached that problem in the best 
scientific tradition, where nothing is taking for granted \cite{Kennefick}. In 1957 we 
find him arguing against any glib analogy to the simpler theory of electrodynamics: 

\

\noindent {\small ``The cardinal feature of electromagnetic radiation is that when radiation 
is produced the radiator loses an amount of energy which is independent of the location 
of the absorbers. With gravitational radiation, on the other hand, we still [in 1957] do 
not know whether a gravitational radiator transmits energy whether there is a near receiver 
or not.'' \cite{DeWitt}}

\

\noindent Two years later, when Bondi and McCrea wrote their paper, the question of the 
reality of gravitational waves was still on their minds. But the electrodynamic analogy 
of their thought experiment is not to electromagnetic radiation, the 
analogy is to inductive energy transport in the near zone such as occurs in a transformer. 
If the receiver is not there the sender simply stores some energy in the magnetic field, 
and gets it back when the AC current is turned off. 

However, our reason for revisiting the Bondi--McCrea example was that we wanted to see 
how the various candidates for a local gravitational energy density fare in a concrete example. 
Equation (\ref{konserv}) makes it clear that each choice of energy density comes with a 
gravitational Poynting vector describing energy transport. Thus, outside matter, 
the energy density proposed by Lynden-Bell and Katz leads to the Poynting vector 

\begin{equation} {\cal S}_i = - \frac{1}{4\pi}\partial_t\Phi \partial_i\Phi 
\end{equation}

\noindent The energy densities preferred by Maxwell and by Bondi lead to the respective 
Poynting vectors 

\begin{equation} {\cal S}_i^\prime = \frac{1}{4\pi}\Phi \partial_i \partial_t\Phi 
\label{Poynting1} \end{equation}

\begin{equation} {\cal S}_i^{\prime \prime} = \frac{1}{8\pi}( \Phi \partial_i\partial_t\Phi 
- \partial_t\Phi\partial_i\Phi ) \ . \label{Poynting2} \end{equation}

\noindent In Section 2 we made the choice between them 
by deciding what boundary conditions to use in order to control the system. But this 
is not at all how the concrete problem is posed. The gravitational potential is controlled from 
inside the system by Tweedledum, and by his agreement with Tweedledee. 

However, among the three, Bondi's Poynting vector ${\cal S}_i^{\prime \prime}$ enjoys 
the advantage that it is divergence free in vacuum. This lends a special significance 
to the flux integral 

\begin{equation} I = \oint_S {\cal S}_i^{\prime \prime} {\rm d}S_i \ , \end{equation}

\noindent where $S$ is a closed surface in vacuum. The surface can be freely deformed 
within the vacuum region without changing the value of the integral. If there is 
vacuum outside the surface it evaluates to zero. There is no net flux out of the surface, 
regardless of how the body inside the surface is changing its multipole moments. But 
if the surface divides two regions containing two distinct bodies it quantifies 
the amount of energy transferred between them \cite{Bondi}. This means that 
Bondi's Poynting vector offers the best way of calculating the energy transferred between 
Tweedledum and Tweedledee. Nevertheless the necessary calculations are very complex. 
They are discussed in Ref. \cite{Tobias}. 

In conclusion, we have found that there is a sense in which Newtonian gravity prefers 
a local energy 
density that is in agreement with that of Lynden--Bell and Katz \cite{LyndenBell}, but 
it is notable that this plays no useful role in the concrete discussion of energy 
transfer between Tweedledee and Tweedledum. There it seems much more helpful to regard 
the energy as localized within the matter. In the end, the work done on the body is 
independent of the way in which gravitational energy is localised \cite{Purdue}. In 
general relativity the question of how to define a useful notion of quasi-local energy 
has given rise to a large literature \cite{Szabados}, while the definition of the total 
energy of an isolated system is clear \cite{Deser}. The warning to relativists is not 
to expect a unique answer, rather we expect several different answers that are useful in 
different ways. The question should be: What quasi-local energy expression is best suited to 
describe energy transport in a given concrete situation? 

\

\

\noindent \underline{Acknowledgement}: It is a pleasure to thank the referees for their 
thoughtful comments.

\

\end{document}